\documentclass[12pt]{iopart}
\usepackage{graphicx} 
\newcommand{\pT}{$p_{T}$}

\begin{document}

\title[Jets and jet-like correlations studies from STAR]{Jet and jet-like correlations studies from STAR }

\author{Helen Caines for the STAR collaboration}

\address{Yale University, New Haven, CT 06520}
\ead{helen.caines@yale.edu}
\begin{abstract}

 I present recent results from jets  and jet-like correlation measurements from STAR. The pp  data are compared to those from Au-Au
collisions to attempt to infer information on the medium produced and how hard scattered partons interact with this matter. Results from d-Au events are utilized to investigate the magnitude of  cold nuclear matter effects on hard scatterings. The evolution of the underlying event  from pp to d-Au collisions is studied.  In heavy-ion collisions, background fluctuations  are the major source of systematic uncertainties in jet measurements. Detailed studies are therefore being made of these fluctuations and recent progress in our understanding  is reported. Jet and jet-hadron correlations results are presented and give clear indications that partonic fragmentation at RHIC is highly modified in the presence of a strongly coupled coloured medium, resulting in a significant broadening and softening of the jet fragments correlation. Finally di-hadron correlations utilizing identified particles as triggers indicate that the ``ridge" is stronger for p+K than for $\pi$ but that the near-side peak per-trigger yield remains unaltered from d-Au to Au-Au collisions.

\end{abstract}


\section{Introduction}

There is now significant evidence from the RHIC experiments, currently being confirmed at the LHC, that the matter created in ultra-relativistic heavy-ion collisions  is strongly coupled with partonic degrees of freedom, the sQGP. Hard probes are being utilized to investigate how partons interact with the medium, and how the medium responds to the traversal of these highly energetic quarks and gluons.  To this end, charged particle di-hadron correlations, full jet reconstruction and particle identified correlations are being studied. These results are compared to similar analyses in pp and d-Au collisions where no QGP is believed to be created.

~

\noindent Before starting to analyze the data, especially for the full jet reconstruction, one needs to ensure that the behaviour of the jet finders is well understood, and that the simulations, from which the corrections are derived are good representations of the data.  The SISCone, Anti-k$_T$ and k$_T$ jet finding algorithms of the FastJet package~\cite{FastJet} were applied to the $\sqrt{s_{NN}}$=200 GeV pp data. The reconstructed raw jet spectra were the same within 10$\%$, confirming that they have similar behaviours in this low multiplicity data. The jet energy resolution was inferred from PYTHIA 6.410~\cite{PYTHIA} simulations tuned to the CDF 1.96 TeV data (Tune A) at the detector level.  Detector level PYTHIA events have been passed through STAR's detector simulation and reconstruction algorithms. The resolution varies from 10-25$\%$, for 40-10 GeV/c jets, being worse for lower jet energies. The dominant source of systematic errors on the corrected jet spectra are the uncertainties in the Barrel Electro-Magnetic calorimeter (BEMC) calibration. This 5$\%$ uncertainty on the individual tower energies translates into an uncertainty on the spectra of up to 40$\%$. The uncertainty on the pp luminosity is an additional 7.7$\%$ and applied separately. 

\section{pp and d-Au}

\subsection{Jet and di-jet cross-sections}

The inclusive jet and di-jet cross-sections  have been measured using the increased statistics of the 2006 data~\cite{DijetXSect}. A  midpoint cone algorithm~\cite{Cone} with a cone radius of 0.7, a split-merge fraction 0.5 and a seed energy of 0.5 GeV was used. When hadronization and underlying event uncertainties are included the data are well described by NLO theory~\cite{NLO1,NLO2}. The measured d-Au jet \pT\ spectrum from the 2008 data set is shown in figure~\ref{Fig:dAuJet}, left panel compared to the binary scaled pp measurement. Within errors the spectra are equivalent, as naively expected for hard processes~\cite{Kapitan}. It should be noted that in this preliminary comparison the anti-k$_{T}$ jet finder was used for the d-Au data whilst  the mid-point cone algorithm was used for the pp data. For this analysis both jet finders were run with a cone radius/jet resolution parameter R=0.4.


\subsection{Cold nuclear matter effects}

\begin{figure}[htb]
	\begin{minipage}{0.46\linewidth}
		\begin{center}
			\includegraphics[width=\linewidth]{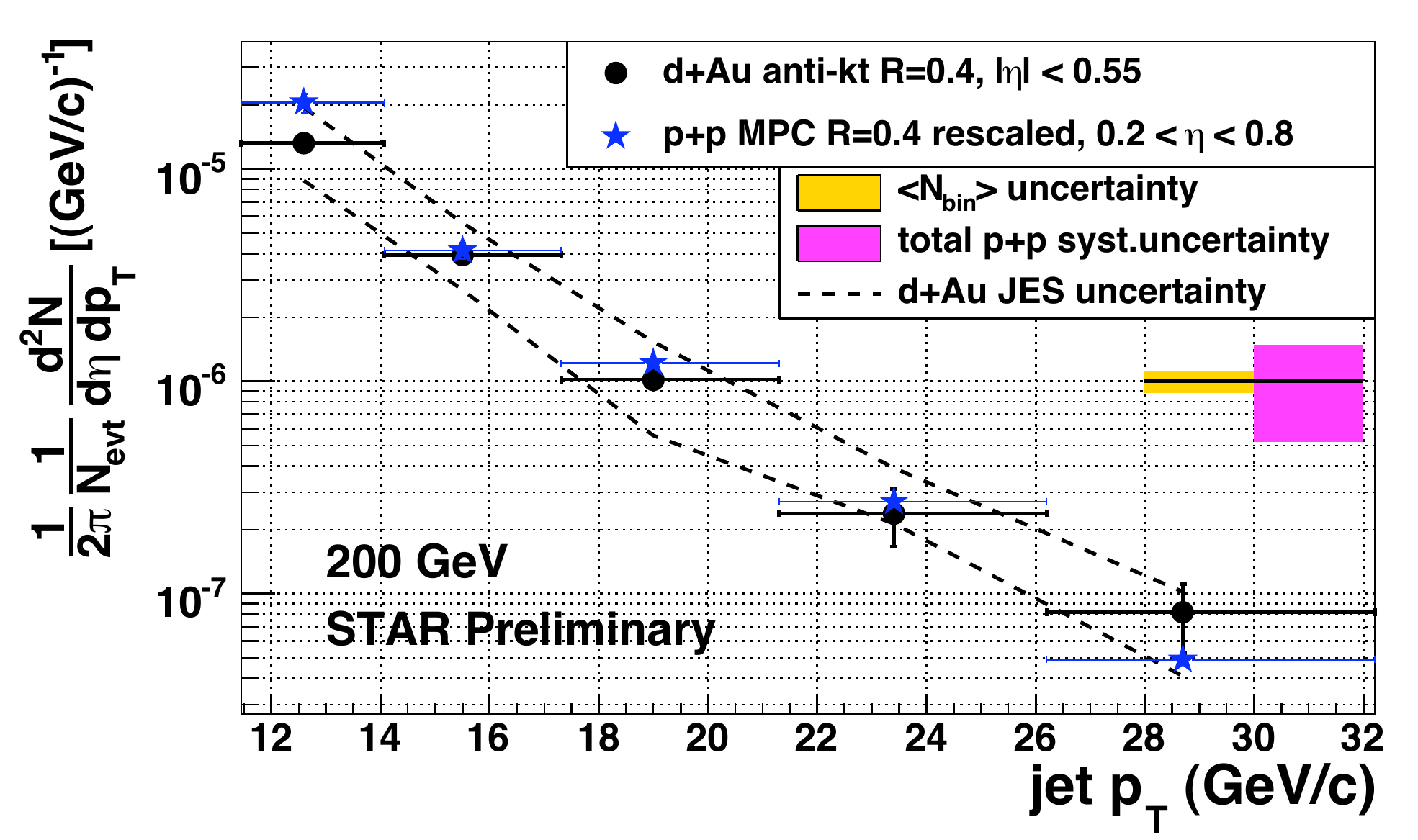}
		\end{center}
	\end{minipage}
	\begin{minipage}{0.46\linewidth}
		\begin{center}
			\includegraphics[width=\linewidth]{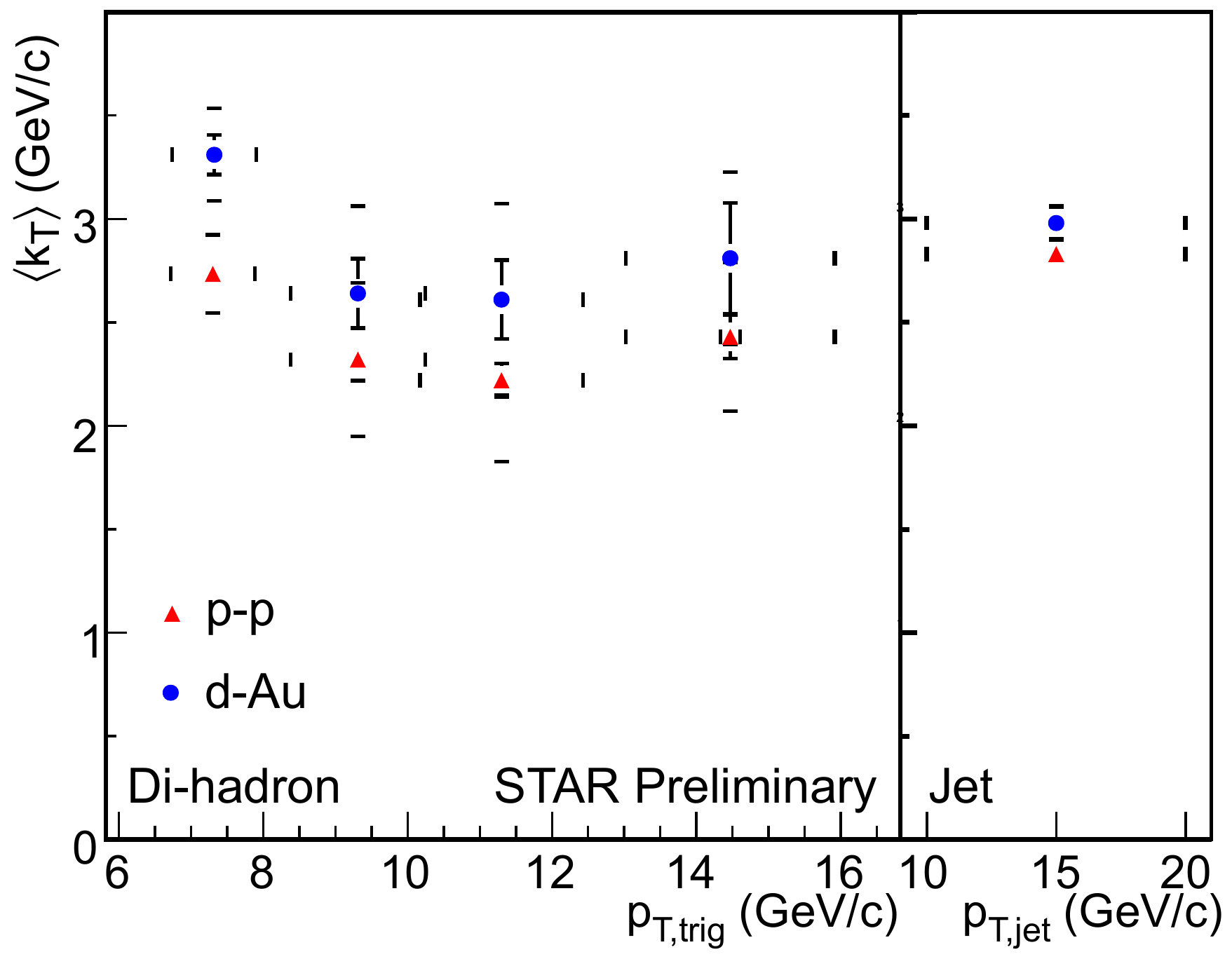}
		\end{center}
	\end{minipage}
	\caption{Left: The corrected \pT\ jet spectra for 0-20$\%$ most central d-Au and binary scaled pp data.  Right: The measured $\langle k_{T} \rangle$  for d-Au and  pp data at $\sqrt{s_{NN}}$=200 GeV from di-hadron correlations and di-jet measurements. Vertical bars show the statistical and systematic uncertainties, horizontal bars indicate the bin widths.}
			\label{Fig:dAuJet}
			\label{Fig:kt}
\end{figure}

Before looking for evidence of partonic scattering/energy loss in Au-Au collisions we investigate jet production in d-Au collisions, where cold nuclear mater effects are expected to be present but no QGP formed.  The presence of the Au nucleus may however induce  additional initial and final state radiation, or result in scatterings of fragmentation particles as they escape the nucleus. These effects may  result  in  more subtle modifications  than that of the overall jet yields. For instance  as changes in the fragmentation patterns or as a broadening of the di-jet $\Delta \phi$ distribution. The presence of such effects are investigated via the variables $\langle  j_{T} \rangle$:   the mean transverse momentum of the fragmentation products with respect to the jet axis, and  $\langle k_{T} \rangle$:  the mean transverse momentum kick given to  di-jet pair. 

~

\noindent The  $\langle j_{T} \rangle$ was measured via di-hadron correlations and found to be constant at $\approx$0.55 GeV/c for all \pT\ triggers  measured and for both pp and d-Au events at  $\sqrt{s_{NN}}$=200 GeV~\cite{Mondal}. Meanwhile the $\langle k_{T} \rangle$, figure~\ref{Fig:kt}:right, is systematically higher for the d-Au data than for the pp data for all \pT\ jet and \pT\ trigger ranges measured. This suggests that cold nuclear matter effects are small but present in the deflection/broadening of partonic trajectories but that fragmentation is unaffected.

\subsection{Underlying events comparisons}

\begin{figure}[htb]
	\begin{minipage}{0.46\linewidth}
		\begin{center}
			\includegraphics[width=0.9\linewidth]{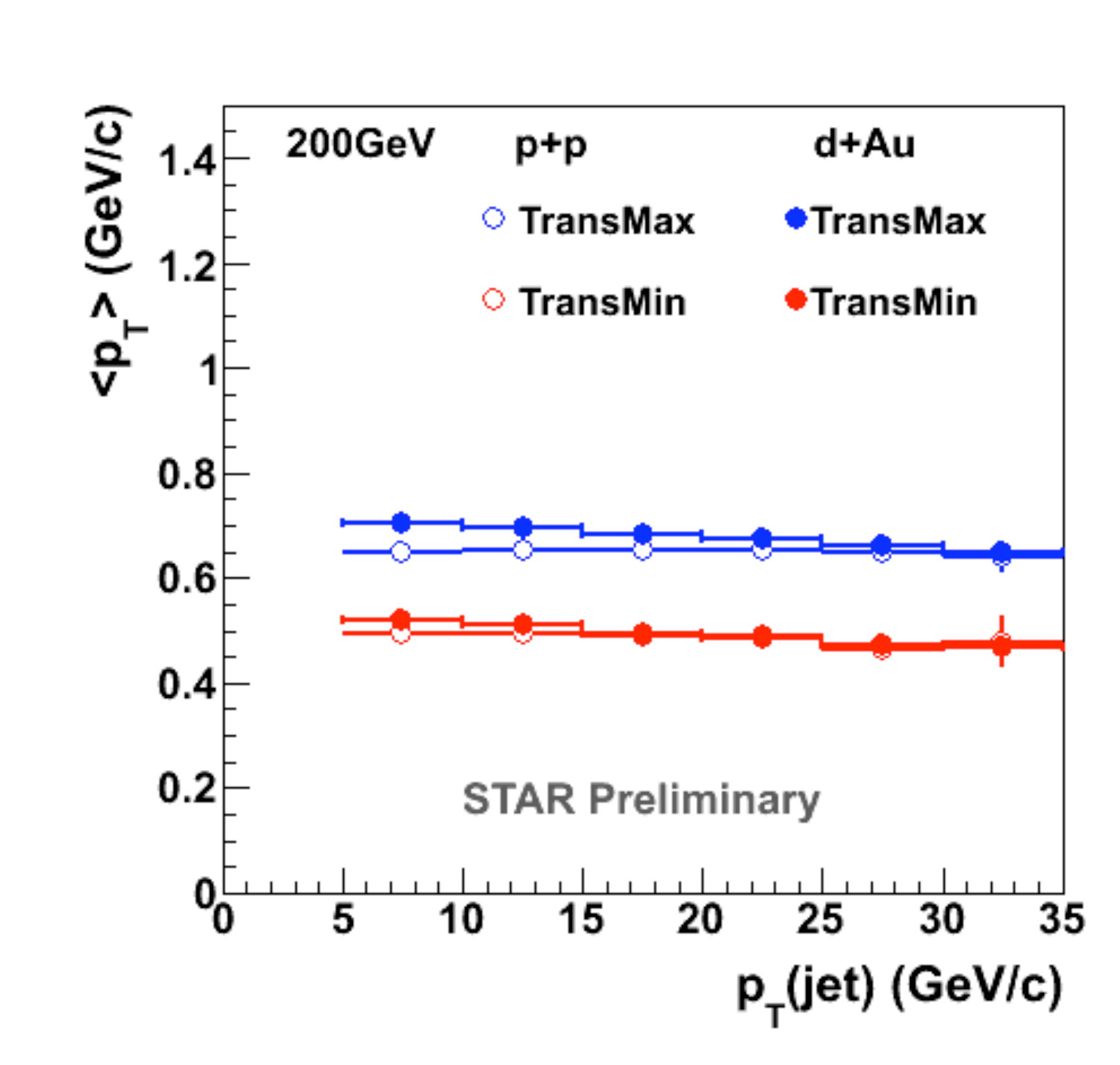}
		\end{center}
	\end{minipage}
	\begin{minipage}{0.46\linewidth}
		\begin{center}
			\includegraphics[width=0.9\linewidth,angle=90]{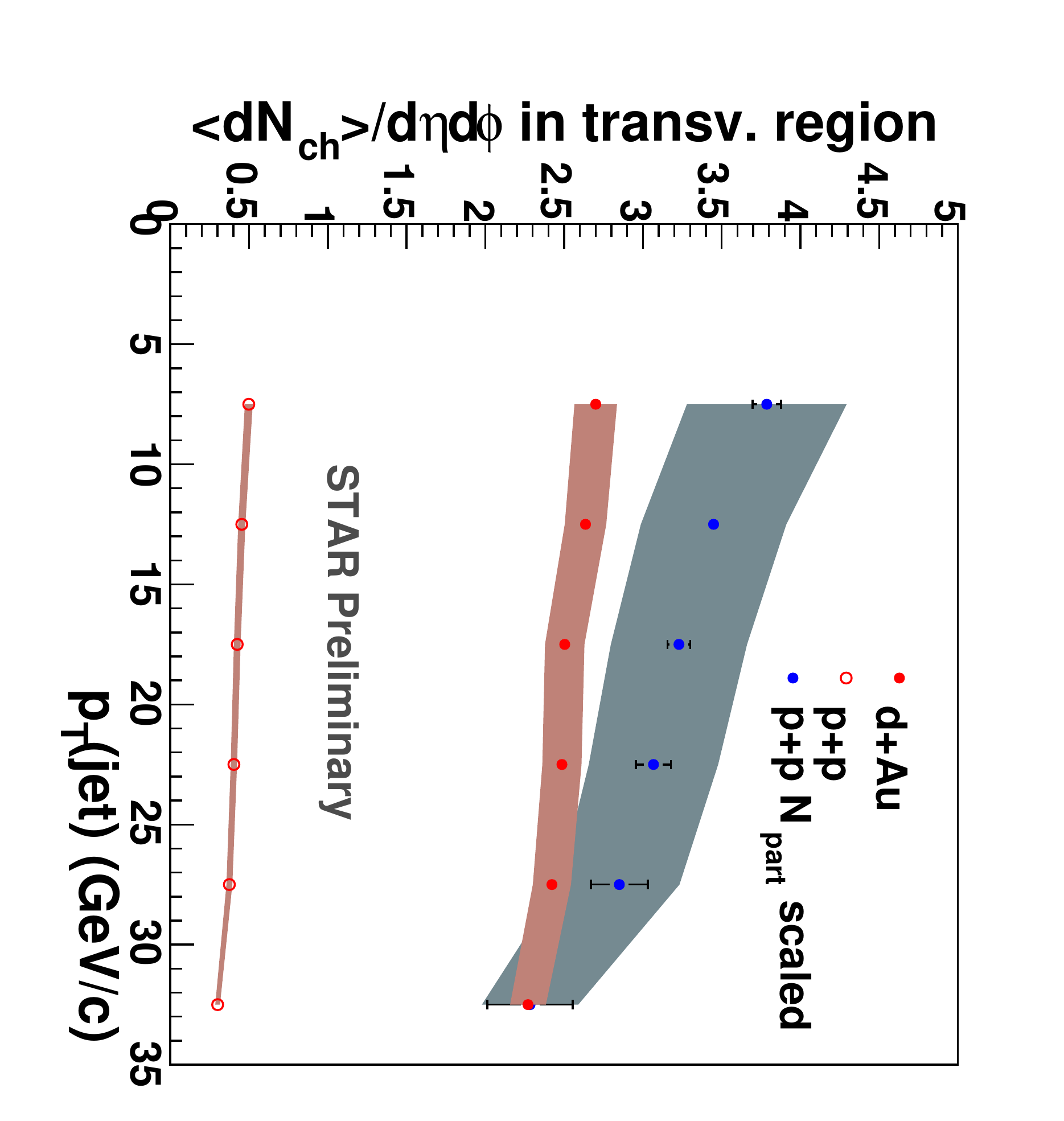}
		\end{center}
	\end{minipage}
	\caption{ Mean p$_{T}$ (left) and mean number of charged particles per unit $\eta$ and $\phi$ (right) in the transverse regions for pp and d-Au collisions.  All data are from  $\sqrt{s_{NN}}$=200 GeV} 
			\label{Fig:UE}
\end{figure}

The underlying event is defined as those particles not produced in the  initial hard scatterings.   In d-Au collisions this underlying event is sizable compared to pp. We investigate the properties of the underlying event in both pp and d-Au events via the study of the transverse regions of the event relative to the reconstructed jet axes. Such an analysis was first undertaken and described  by CDF~\cite{CDF}.  As per CDF, the transverse sector containing the largest charged particle multiplicity  is  called the TransMax region, and the other is  termed the TransMin region. The mean transverse momentum in both sectors is  similar for pp and d-Au events, left panel of figure~\ref{Fig:UE}. The average number of charged particles per unit $\eta$ and $\phi$ however increases by $\sim$factor 5 from pp and d-Au collisions,  right panel of figure~\ref{Fig:UE}. This increase in particle production is slightly less than $N_{part}$ scaling of the pp data would predict, also shown in figure~\ref{Fig:UE}. More d-Au underlying event studies were presented in \cite{Bielcikova}.
 
\section{Au-Au}
\subsection{Studies of the background fluctuations}
\begin{figure}[htb]
	\begin{minipage}{0.46\linewidth}
		\begin{center}
			\includegraphics[width=0.9\linewidth]{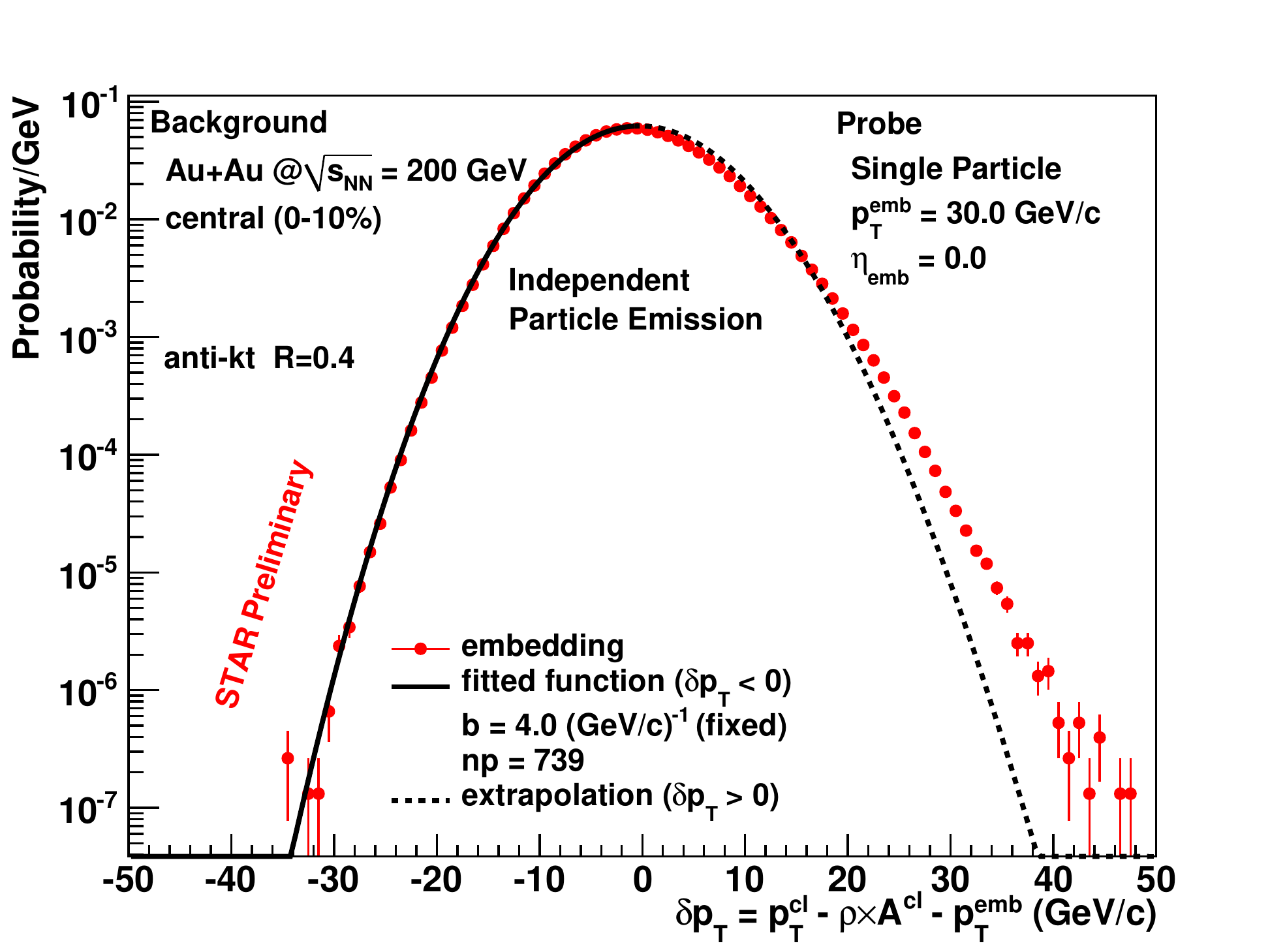}
		\end{center}
	\end{minipage}
	\begin{minipage}{0.46\linewidth}
		\begin{center}
			\includegraphics[width=0.9\linewidth]{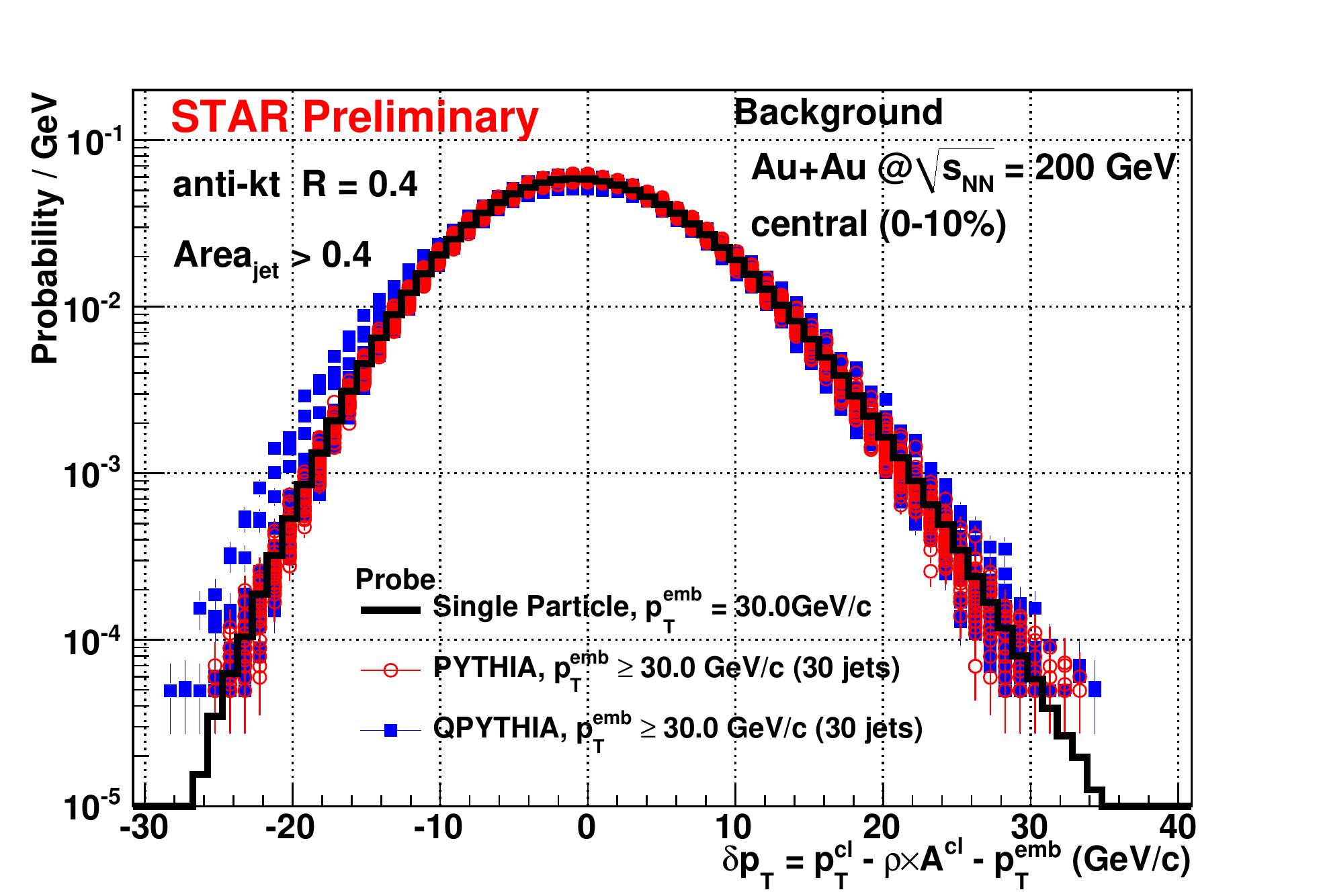}
		\end{center}
	\end{minipage}
	\caption{Left: The $\delta p_{T}$ distribution for 30 GeV/c single particle probes. Right:  The $\delta p_{T}$ distributions for 30 GeV/c probes:  black - single  particles, red -  PYTHIA jets  and blue  qPYTHIA  jets.  All data are for 0-10$\%$  $\sqrt{s_{NN}}$=200 GeV collisions.}
			\label{Fig:Deltapt}
\end{figure}

In Au-Au collisions the underlying event dominates the particle production even in events containing a hard scattering. This background must be studied and subtracted properly from any jet measurements before detailed conclusions can be drawn. To this end detailed studies of the underlying event and its fluctuations in Au-Au events are being performed by STAR.  Schematically the measured jet \pT\ spectrum is $ \frac{d \sigma_{AA}}{dp_{T}} \propto \frac{d \sigma_{pp}}{dp_{T}} + \rho \times A \pm F(A,p_{T}) $, where $\rho \times A $ is the average event-by-event energy in an area A, and F(A, \pT) are the background fluctuations,  that may or may not be correlated within the event. If particle emission is independent then the particle multiplicity fluctuations should be well described by a Poissonian distribution and the $\langle p_{T} \rangle$ fluctuations, for a fixed multiplicity, by a Gamma distribution~\cite{Tannenbaum}. In this case $F(A,p_{T}) =  Poisson(N(A)) \otimes F_\Gamma (N(A),\langle p_{T} \rangle )$, where N(A) is the number of particles in a cone area A. To study the fluctuations in the data we utilize general probe embedding (GPE). In this data driven approach a  probe of a known momentum is embedded into a real Au-Au event. The jet finder is run and the jet containing the probe identified. This is repeated for many probes and events and the $\delta p_{T}$ distribution evaluated, where $\delta p_{T} = p_{T,jet}^{rec} - \rho \times A - p_{T,probe}$. In this way the fluctuations in the event can be mapped over several orders of magnitude.  $\delta p_{T}$ distributions for 0-10$\%$ most central data are shown in figure~\ref{Fig:Deltapt}. On the left  is the $\delta p_{T}$  calculated for single 30 GeV/c probe particles. The solid line is a fit to a Gamma distribution for $\delta p_{T} < $ 0. It can be seen that the fit gives a good description of the LHS of the data but underestimates the RHS, as it should since this part of the distribution contains the true jet signal. The fit is consistent with the background clusters being formed from $\sim$370 independent ``sources" with an exponentially falling spectra with a fixed  $\langle p_{T} \rangle$= 0.5 GeV/c, in reasonable agreement to the RHIC ``bulk" data results. The right plot shows the 
$\delta p_{T}$ distribution for single particle probes in black, PYTHIA jet probes in red and qPYTHIA~\cite{qPythia}, ($\hat{q}$=5GeV$^2$/fm),  jet probes in blue. It can be seen that the distributions are in reasonable agreement, indicating that the background fluctuations under a jet are  approximately independent of the jet fragmentation. More details can be found here \cite{Deltapt}. 

\subsection{Evidence for jet broadening and softening in Au-Au collisions}

 Full jet reconstruction measurements at STAR have already given direct evidence of softening and broadening of the jet fragmentation in Au-Au collisions. First the jet R$_{AA}$, while higher than that of single particles remains less than unity for a R=0.4. Second the R=0.2/R=0.4 ratio of reconstructed jets is  less for Au-Au data than for pp~\cite{QM09}. Taken together these results indicate that  the algorithms  recover less of the original partonic energy compared to the same code operating on pp data, and that this is due to  particles being emitted at larger cone angles in Au-Au collisions than in pp. Evidence that the energy loss has a path-length dependence comes from the measured di-jet coincidence rate. In this analysis ``trigger" jets are identified that contain a barrel electro-magnetic calorimeter tower with E$_{tow}>$ 5.4 GeV/c and to have a reconstructed jet \pT\ $>$ 20 GeV/c when only particles with \pT\ $>$ 2 GeV/c are considered by the Anti-k$_{T}$ jet finder. This high z fragmentation requirement biases the ``trigger" jet to being preferentially emitted from the surface of the medium and/or to have only minimally interacted with the medium. This ``trigger" jet surface bias in turn maximizes the distance traversed by the recoil jet through the medium. If partonic energy loss is dependent on the path-length, the suppression of the recoil jets should be greater than that observed for the unbiased jet R$_{AA}$ measure.  When the relative probability of reconstructing a di-jet pair in Au-Au is compared to that in pp it is indeed found to be suppressed by an approximate  factor of 5~\cite{QM09di}, i.e. a much stronger rate of suppression than observed for the inclusive jets.

 \subsection{Jet-hadron correlations}
 
\begin{figure}[htb]
	\begin{minipage}{0.46\linewidth}
		\begin{center}
		\includegraphics[height=0.9\linewidth,angle=90]{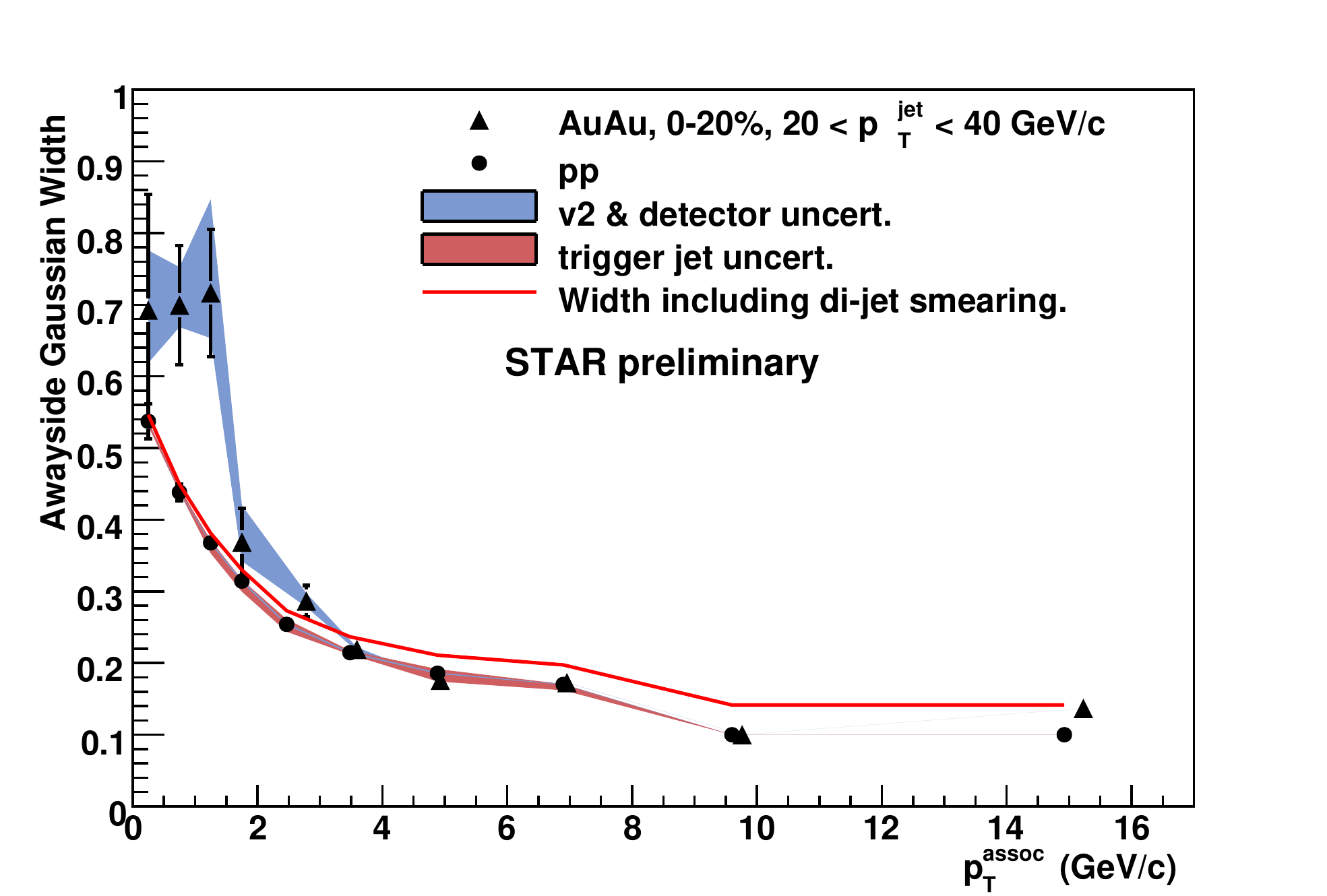}
		\end{center}
	\end{minipage}
	\begin{minipage}{0.46\linewidth}
		\begin{center}
		\includegraphics[width=0.9\linewidth]{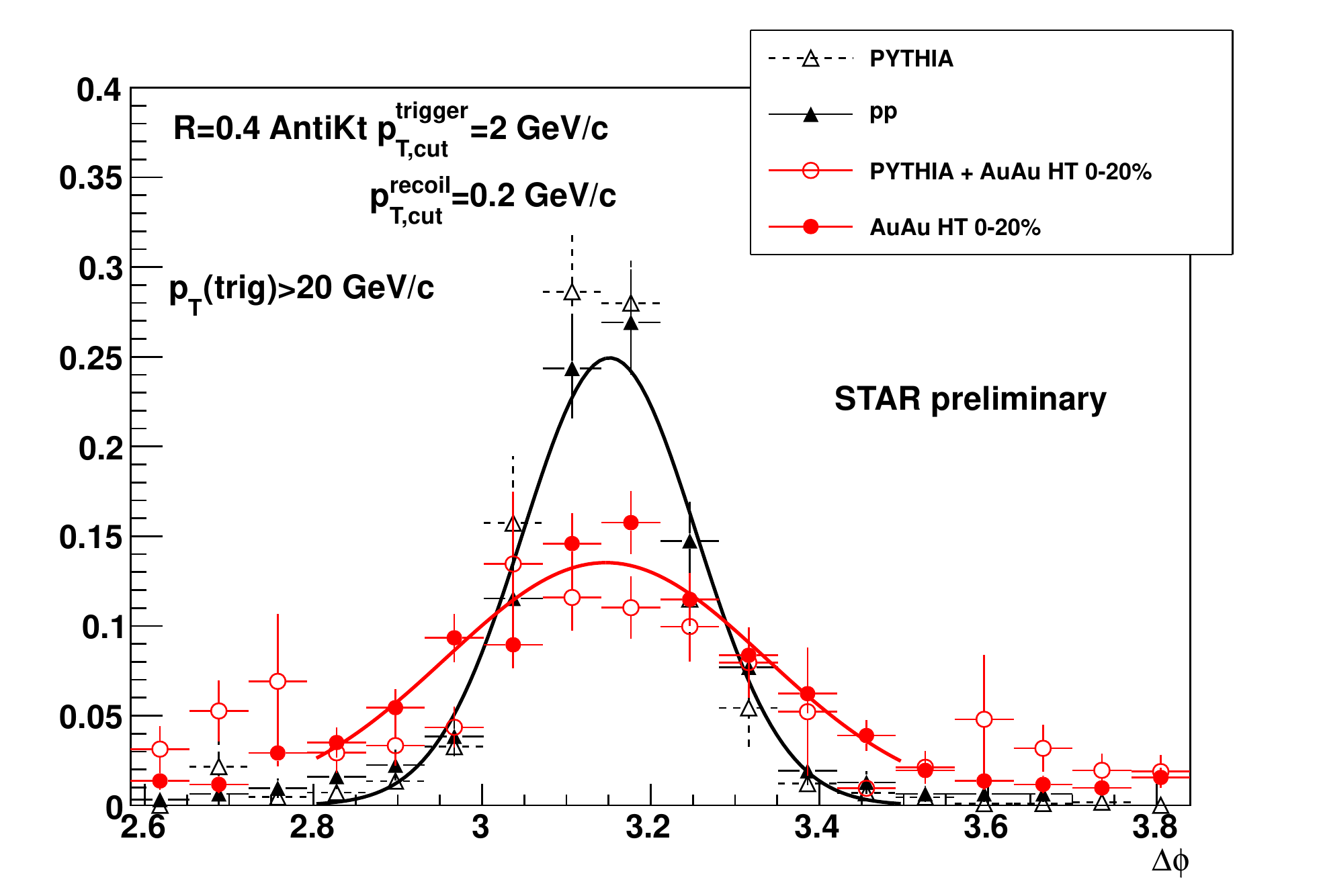}
		\end{center}
	\end{minipage}
	\caption{Left: The Gaussian widths of the away-side correlations as a function of p$_{T}^{assoc}$. Right: Di-jet $\Delta \phi$ distributions for various data sets at $\sqrt{s_{NN}}$=200 GeV collisions. Solid curves are Gaussian fits to the data.}
			\label{Fig:deltaPhi}
\end{figure}

\begin{figure}[htb]
	\begin{minipage}{0.46\linewidth}
		\begin{center}
			\includegraphics[width=0.9\linewidth]{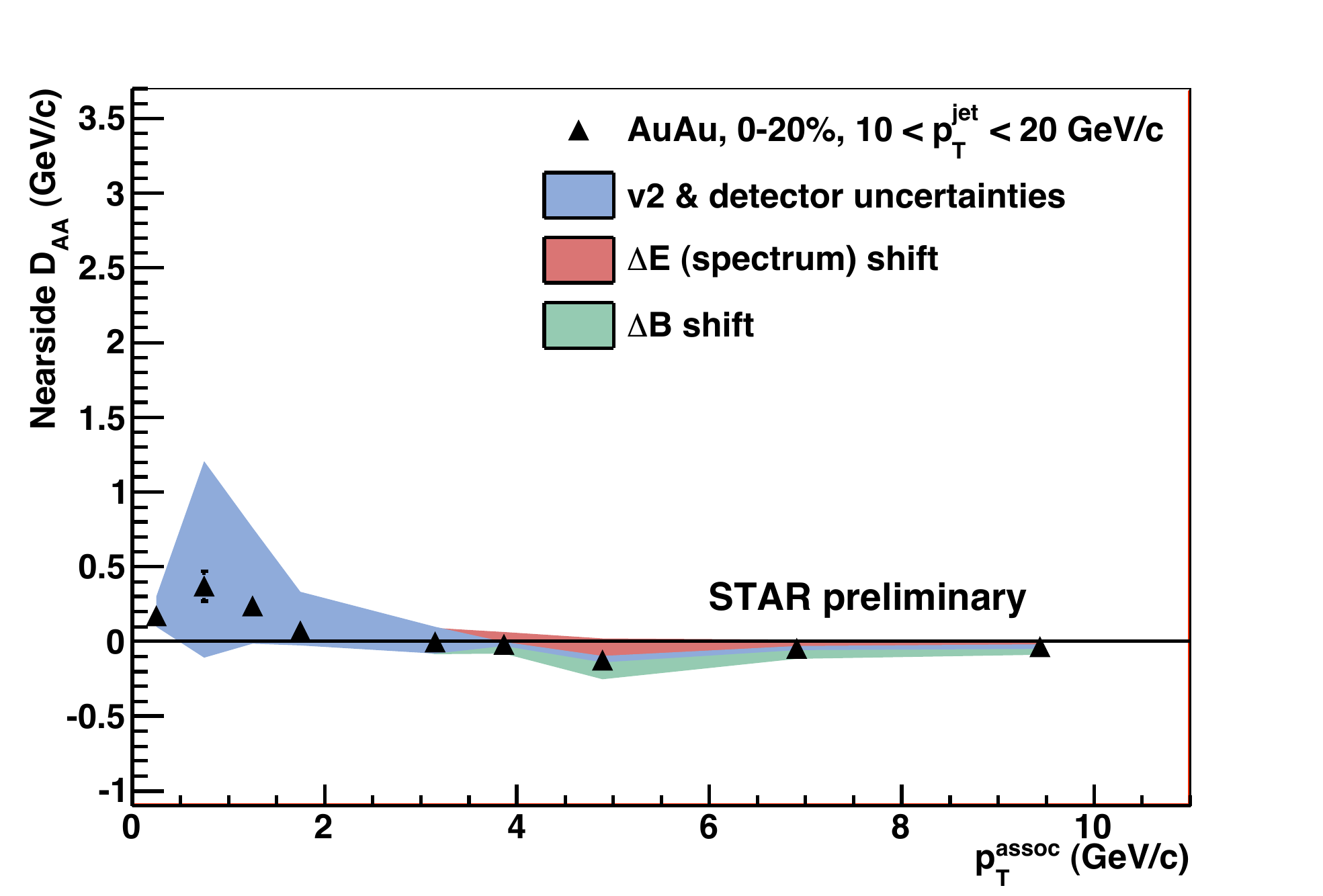}
		\end{center}
	\end{minipage}
	\begin{minipage}{0.46\linewidth}
		\begin{center}
			\includegraphics[width=0.9\linewidth]{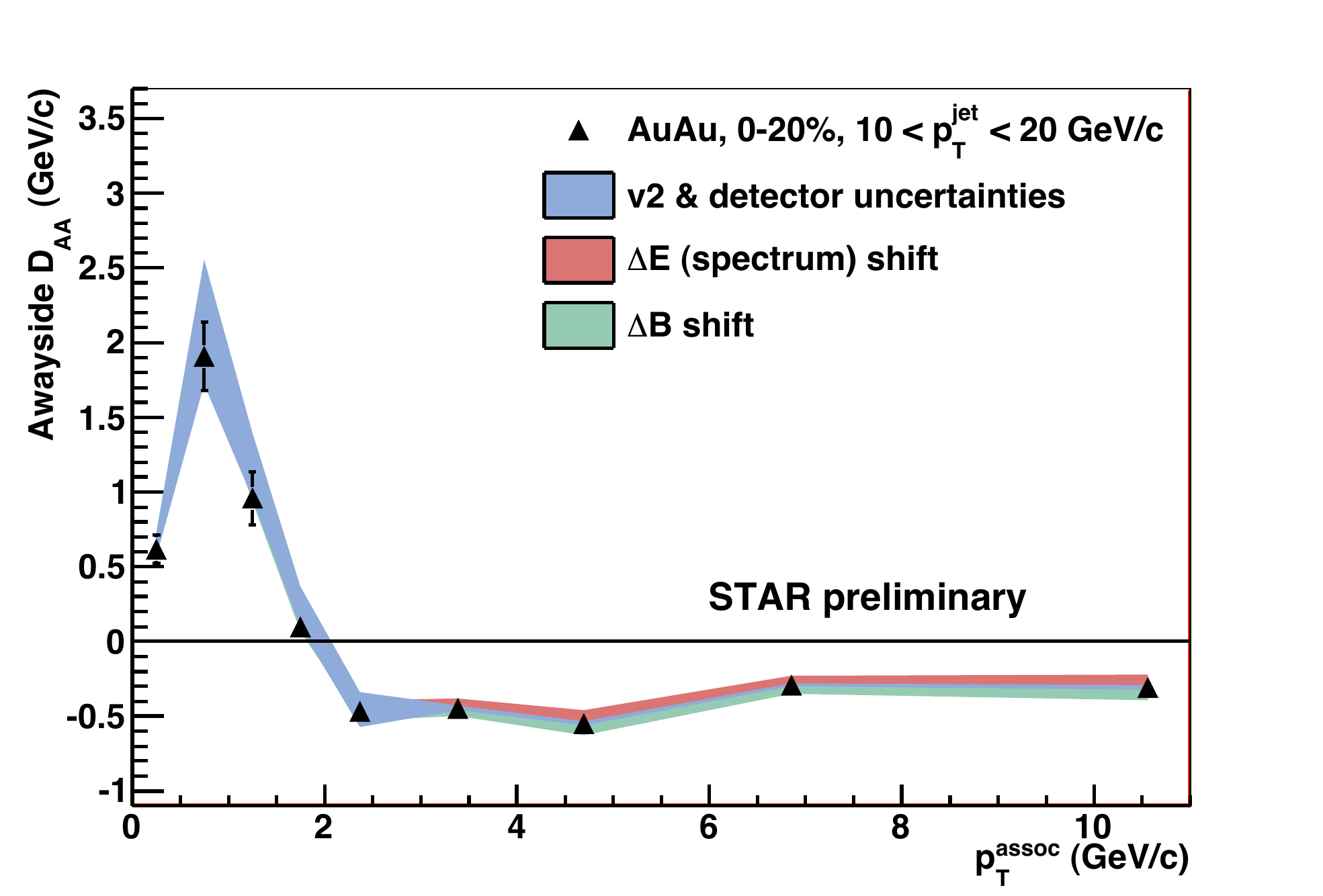}
		\end{center}
	\end{minipage}
	\caption{Jet-hadron D$_{AA}$ distributions for  the near-side (left)  and away-side (right). For details on the systematic uncertainty bands see \cite{Ohlson}.}
			\label{Fig:DAA}
\end{figure}

 To investigate this jet broadening and softening further we turn to jet-hadron correlations. In this analysis we use a ``trigger" jet (defined as in the di-jet analysis above) to determine the jet axis and examine the $\Delta \phi$ correlation of  {\it all} charged particles in the event relative to this axis, for more details on this analysis see \cite{Ohlson}. If the per trigger $\Delta \phi$ distributions for pp and Au-Au event are plotted as a function of the associated charged particle \pT\ striking differences, again reflecting the softening and broadening of the jet fragments,  can be seen already in the raw  away-side jet-hadron correlations~\cite{OhlsonWW}. These results are summarized in figures~\ref{Fig:deltaPhi}  and \ref{Fig:DAA}. Figure~\ref{Fig:deltaPhi}  shows the Gaussian width of the away-side  correlation in pp and Au-Au. For low \pT\ associated particles the width of the Au-Au distributions is much broader, a complementary increase in the low \pT\ yields is also observed~\cite{Ohlson}. However, this broadening could be caused by re-scattering of the initial parton, rather than a modification of the fragmentation. We therefore looked at the $\Delta \phi$ distribution of identified di-jet events in pp and Au-Au data, PYTHIA events, and PYTHIA jets embedded into Au-Au events. The results are shown in right plot of figure~\ref{Fig:deltaPhi}. As can be seen the distribution is broader for the Au-Au data, however much of this broadening can be attributed to de-resolution of the jet axis due to the large underlying event, as demonstrated  by the PYHTIA+Au-Au event histogram. The red curve in the left plot of figure~\ref{Fig:deltaPhi} indicates the expected  width of the away-side  $\Delta \phi$ distribution if the Au-Au fragmentation was pp-like, but with the jet axis direction smeared to reproduced the width of the $\Delta \phi$  Au-Au di-jet data. Clearly such a smearing cannot fully explain the observed broadening, and it also does not explain the enhanced low \pT\ yields.
 
The difference in integrated yields (D$_{AA} = Yield_{AA}\times\langle p_{T}^{assoc}\rangle-Yield_{pp}\times\langle p_{T}^{assoc}\rangle$) of the near- and away-side correlations as a function of p$_{T}^{assoc}$ are plotted in figure~\ref{Fig:DAA}. As expected due to the ``surface" bias of the trigger, the near-side D$_{AA}$ is consistent with zero for all  p$_{T}^{assoc}$. 
This means that there is an approximate energy balance, and a similarity of the associated \pT\  particle distributions for Au-Au and pp data for the trigger jet.
While the away-side data indicate that the low \pT\ hadron enhancement in the Au-Au data is approximately matched by a high \pT\ associated particle suppression. This  suggests that the broadening and softening observed in the away-side correlation data is indeed due to a modification of the partonic fragmentation and not from residual soft  background particles.

\subsection{PID di-hadron correlations}

\begin{figure}[htb]
	\begin{minipage}{0.46\linewidth}
		\begin{center}
			\includegraphics[width=0.9\linewidth]{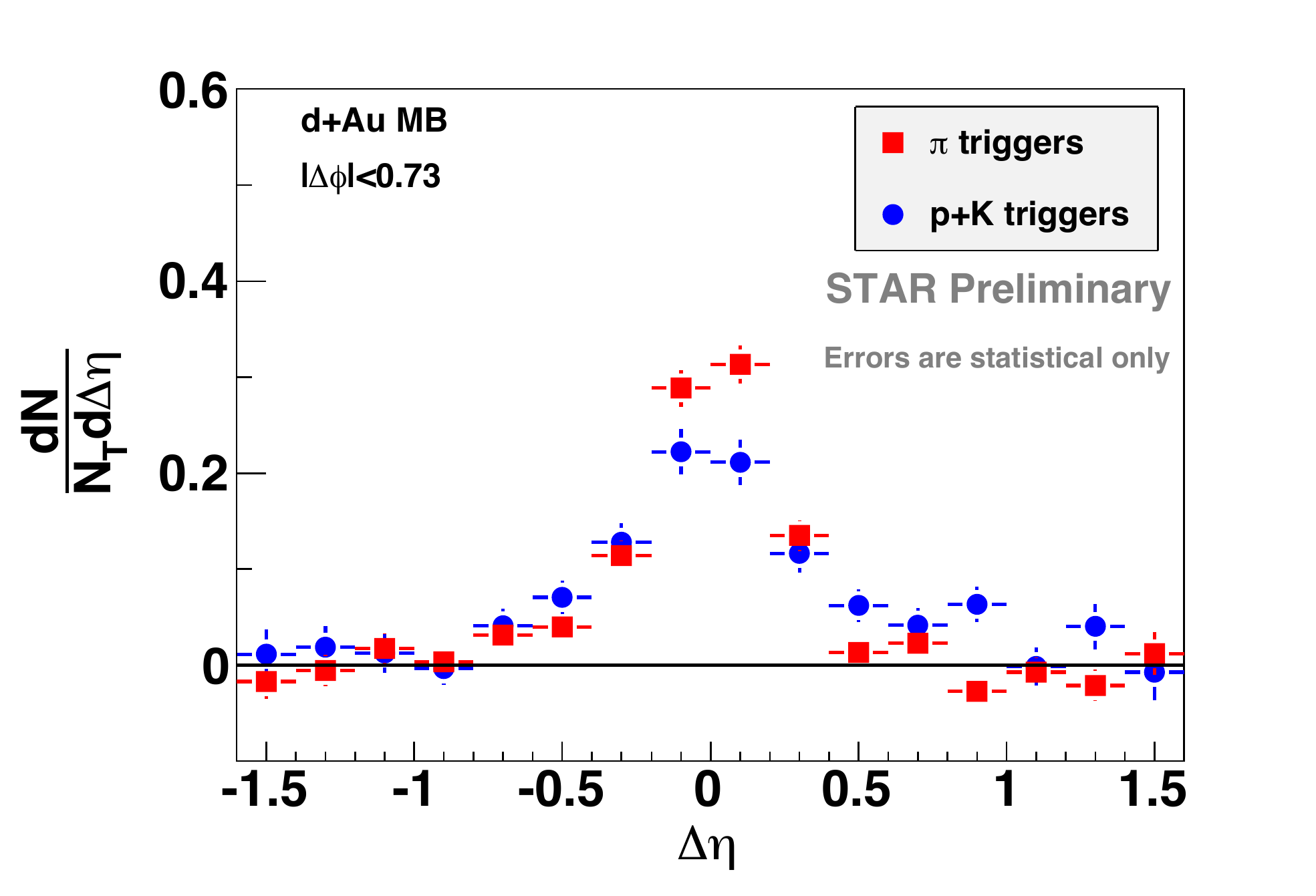}
		\end{center}
	\end{minipage}
	\begin{minipage}{0.46\linewidth}
		\begin{center}
			\includegraphics[width=0.9\linewidth]{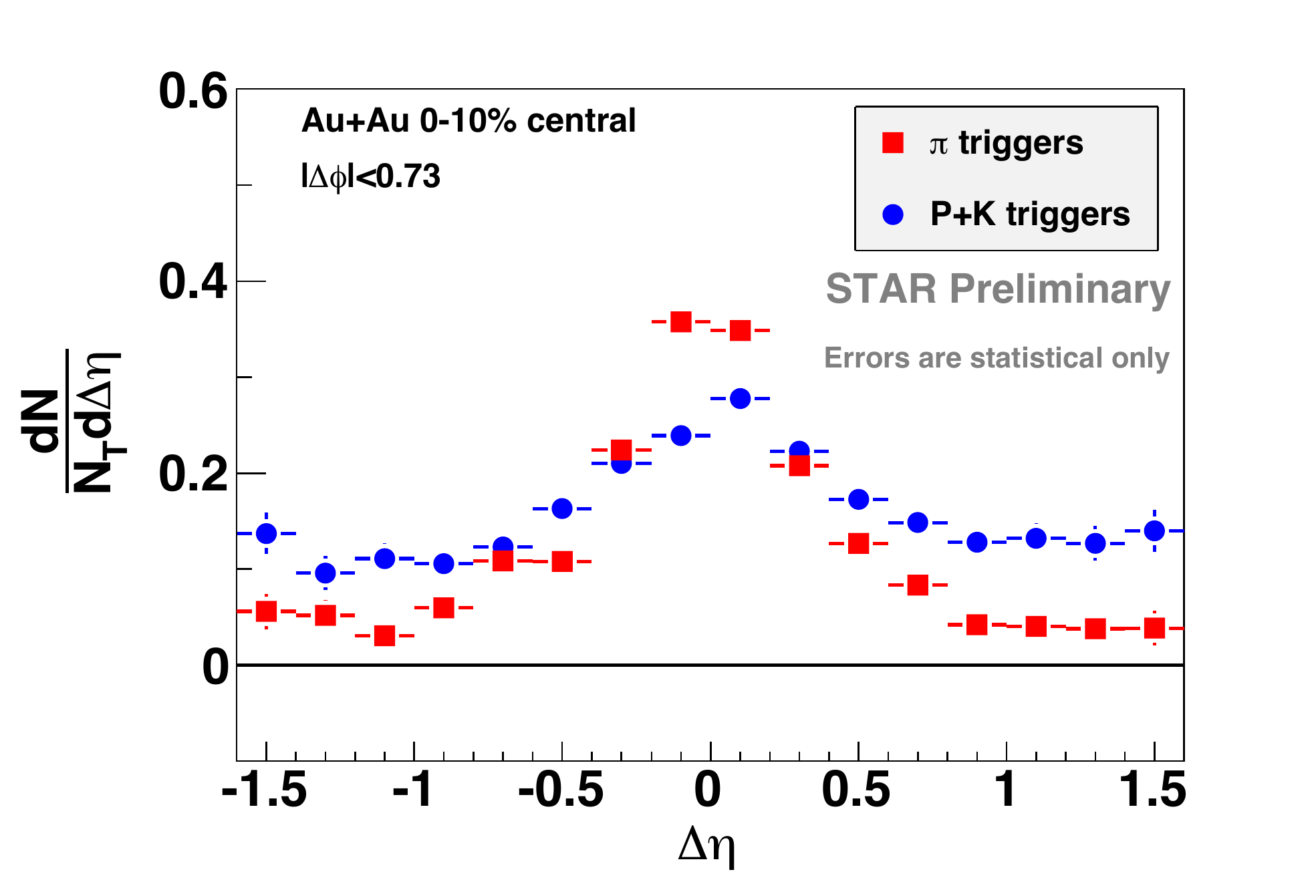}
		\end{center}
	\end{minipage}
	\caption{p+K and $\pi$ triggered di-hadron near-side $\Delta \eta$ correlations. Left:  d-Au collisions. Right: Au-Au collisions.}
			\label{Fig:PID}
\end{figure}

The  particle type dependence of di-hadron correlations should give us further insight into the partonic energy loss. For the studies reported below the relativistic rise of the specific energy loss in the TPC is utilized to statistically separate p+K from $\pi$. For both the d-Au and Au-Au analyses trigger particles with \pT\ =4-6 GeV/c were correlated with p$_{T}^{assoc} >$ 1.5 GeV/c.  The near-side $\Delta \eta$ projections, figure~\ref{Fig:PID}, show that the $\pi$-triggered data result in higher per trigger yields on the near-side than p+K-triggered data in both  d-Au and Au-Au collisions~\cite{Kauder}. The Au-Au data sits on a pedestal, the phenomenon traditionally known as {\it ``the ridge"}, with a higher pedestal  for the p+K-triggered correlations that for the  $\pi$-triggered data. This suggests that  protons and/or kaons may have a higher v$_{3}$ component than the $\pi$, see \cite{Sorensen} for more details on flow and fluctuation measures from STAR. The intermediate \pT\ trigger range selected for this analysis is  in the  region often successfully described by ``recombination" models~\cite{Reco}. However, if the production of protons were dominated by  recombination of quarks from the ``bulk", there should be no correlated particle production with these baryons. In the na\"{\i}ve recombination scheme this would result in a reduced per trigger yield for protons in Au-Au di-hadron correlations.  No such  reduction, compared to d-Au data,  is observed in figure~\ref{Fig:PID}.

\section{Summary and Outlook}

In summary, full jet reconstruction has shown that the pp single inclusive and di-jet cross-sections are well described by NLO calculations when hadronization and the underlying event are taken into the account. The underlying event multiplicity shows approximate N$_{part}$ scaling from pp to d-Au events, while the underlying event  $\langle p_{T} \rangle$ stays constant within errors. Meanwhile the jet cross-section is consistent, within errors, with N$_{bin}$ scaling. Jet fragmentation appears unaffected by cold nuclear matter effects but there is a small, but systematical, increase in the measured k$_{T}$.  Our understanding of the Au-Au background fluctuations is improving rapidly, which will lead to a significant improvement in our  jet analyses.
~
\noindent Jet-hadron collisions have been utilized to minimize the effects of background fluctuations on our results. They indicate that the quenched high \pT\ fragmentation products reemerge as numerous low \pT\ particles at large angles with respect to the jet axis. These results reinforce those reported at QM09.

~

\noindent Finally,  identified trigger particle di-hadron correlations do not show the strong suppression of the near-side per trigger yield in Au-Au events. Such as dilution would be expected from na\"{\i}ve recombination models due to ``non-jet" triggers  dominating the baryon analyses. The p+K triggered correlations sit on a significantly increased plateau compared to those from $\pi$ triggers, when viewed in $\Delta \eta$ projections, suggesting that the protons and/or kaons experience the effects of ``triangular flow" more severely than pions.

\ack 
The author wishes to acknowledge the contributions of the Bulldog High Performance computing facility of Yale University to this work, and the support of the DoE.

\end{document}